\documentclass[prd,twocolumn,showpacs,preprintnumbers,floats,floatfix,
nofootinbib]{revtex4}
\usepackage{graphicx}
\usepackage{dcolumn}
\usepackage{bm}
\usepackage{amssymb}
\def\spose#1{\hbox to 0pt{#1\hss}}
\def\lta{\mathrel{\spose{\lower 3pt\hbox{$\mathchar"218$}}
     \raise 2.0pt\hbox{$\mathchar"13C$}}}
\def\gta{\mathrel{\spose{\lower 3pt\hbox{$\mathchar"218$}}
     \raise 2.0pt\hbox{$\mathchar"13E$}}}
\newcommand{\be}{\begin{equation}}
\newcommand{\en}{\end{equation}}
\newcommand{\bea}{\begin{eqnarray}}
\newcommand{\ena}{\end{eqnarray}}

\def\setR{\mathbb{R}}
\def\setV{\mathbb{V}}
\def\setS{\mathbb{S}}
\def\setH{\mathbb{H}}

\def\setD{\mathbb{D}}

\begin{document}

\title{Green Functions for Topology Change}

\author{J\'er\^ome Martin}
\email{jmartin@iap.fr}
\affiliation{Institut d'Astrophysique de Paris-GReCO, 98 bis 
Boulevard Arago, 75014 Paris, France}

\author{Nelson Pinto-Neto} 
\email{nelson@cbpf.br}
\affiliation{Laboratorio de Cosmologia e Fisica Experimental de Altas
Energias, Centro Brasileiro de Pesquisas Fisicas, Rua Dr. Xavier
Sigaud 150, Urca, Rio de Janeiro CEP 22290-180-RJ, Brazil}

\author{Ivano Dami\~ao Soares}
\email{ivano@cbpf.br}
\affiliation{Laboratorio de Cosmologia e Fisica Experimental de Altas
Energias, Centro Brasileiro de Pesquisas Fisicas, Rua Dr. Xavier
Sigaud 150, Urca, Rio de Janeiro CEP 22290-180-RJ, Brazil}

\date{\today}

\begin{abstract} 
We explicitly calculate the Green functions describing quantum changes
of topology in Friedman-Lema\^{\i}tre-Robertson-Walker Universes whose
spacelike sections are compact but endowed with distinct
topologies. The calculations are performed using the long wavelength
approximation at second order in the gradient expansion. We argue that
complex metrics are necessary in order to obtain a non-vanishing Green
functions and interpret this fact as demonstrating that a quantum
topology change can be viewed as a quantum tunneling effect. We
demonstrate that quantum topological transitions between curved
hypersurfaces are allowed whereas no transition to or from a flat
section is possible, establishing thus a selection rule. We also show
that the quantum topology changes in the direction of negatively
curved hypersurfaces are strongly enhanced as time goes on, while
transitions in the opposite direction are suppressed
\end{abstract}

\maketitle

\section{Introduction}
Combining the principles of General Relativity with those of Quantum
Mechanics seems necessary in order to understand more completely the
physical nature of the gravitational field. As a result, new
properties are expected to emerge. For instance, in classical General
Relativity, topology changes are forbidden in the sense that their
presence would necessarily imply the appearance of either
singularities or closed timelike curves, a result known as the Geroch
theorem, see Ref.~\cite{Ge}. On the other hand, topology changes are
widely believed to become allowed if the quantum-mechanical nature of
the gravitational field is taken into account. Studying this problem
is notoriously known as a difficult technical task and in the
literature only very general aspects of the issue have been discussed
so far, see e.g. Refs.~\cite{Horo,CM}. Recently, a model where the
theoretical ideas about topology change can be implemented concretely,
at the level of equations, has been proposed in Ref.~\cite{LMPS}. The
metric of this model is given by a
Friedmann-Lema\^{\i}tre-Robertson-Walker-like metric where the
curvature of the spacelike sections, usually denoted as ${\cal K}$, is
now a time dependent function, ${\cal K}={\cal K}(t)$
\begin{equation}
\label{basicmetric}
{\rm d}s^2=-N^2(t){\rm d}t^2+a^2(t)\biggl\{{\rm d}\xi ^2+\biggl[
\frac{\sin (\sqrt{{\cal K}}\xi )}{\sqrt{{\cal K}}}\biggr]^2
{\rm d}\Omega ^2(\theta ,\varphi )\biggl\}.
\end{equation}
The three-dimensional spacelike hypersurfaces $\setV ^3$ of the model
may be endowed with a large class of topologies compatible with such
geometry. Technically, the metric has always the form displayed above
(since it is locally defined) but the ranges of variations of the
coordinates are modified. The scalar curvature of the metric is always
constant (for fixed $t$) and equal to $6{\cal K}/a^2$. We will
restrict our considerations to compact (and orientable) spaces in
order to avoid possible surface terms in the Hamiltonian formalism. We
can show that any compact three manifold with constant curvature is
homeomorphic to ${\setV}^3/{\Gamma}$, where the universal covering
${\setV}^3$ is either $\setR^3$, $\setS^3$ or $\setH^3$, according to
the sign of ${\cal K}$ (${\cal K}=0$, ${\cal K}<0$, ${\cal K}>0$,
respectively). The group $\Gamma$ is the group of covering
transformations. In three dimensions the closed (i.e. compact without
boundary) space is now homeomorphic to a polyhedron the faces of which
are identified by pairs, for a general review see Ref.~\cite{LRL}.

\par 

Let us first consider the case ${\cal K}>0$. The universal covering is
$\setS^3$. Since it is compact, all the three-dimensional spaces
admitting this universal covering are also compact. There is an
infinite number of such spaces. We will only consider two cases:
$\setS^3$ itself and the Poincar\'e dodecahedral space, $\setD^3\equiv
\setS^3/I^*$, where $I^*$ is the binary symmetry group of the
icosahedron. The ranges of variation of the coordinates
($\xi$,$\theta$,$\varphi$) can always be written as
\begin{equation}
\label{variables}
0\leq\xi\leq \frac{\Xi (\theta,\varphi;\setV^3)}{\sqrt{{\cal K}}}\, ,
\quad 0\leq\theta\leq\pi \,  \quad 0\leq\varphi\leq\pi.
\end{equation}
The symbol $\setV^3$ in the argument of the function $\Xi $ indicates
that $\Xi(\theta,\varphi;\setV^3)$ is not the same function for
different $\setV^3$ ($\setS^3$ or $\setD^3$). Of course, the case
$\setS^3$ is very simple and we just have $\Xi
(\theta,\varphi;\setS^3)={\pi}$. This guarantees that no conical
singularities will appear in this case. If $\setV^3=\setD^3$ the
function $\Xi$ is also not arbitrary but its explicit expression is
much more complicated. For the case ${\cal K}<0$, where the universal
covering is $\setH^3$, the classification of closed three-dimensional
spaces is still an open question, but it is known that the volumes of
the polyhedron are fixed as in the ${\cal K}>0$ case. Finally, the
case ${\cal K}=0$ have six different tessellations by polyhedrons,
some of them with arbitrary volume for their polyhedrons (for more
details see again Ref.~\cite{LRL}). What is important for us here is
that the topologies compatible with one ${\cal K}$ are not compatible
with other ${\cal K}$ of distinct sign. Hence, a change in the sign of
${\cal K}$, or in the sign of the curvature scalar $6{\cal K}/a^2$,
necessarily indicates a change of topology. Our strategy here is to
investigate whether this change is possible quantum mechanically.

\par

The quantization of this model was carried out in Ref.~\cite{LMPS}. In
that article, it was shown that a Hamiltonian formulation necessarily
requires a passage to a {\em midisuperspace} description. It is not
possible to construct a {\em minisuperspace} Hamiltonian from the
metric Eq.~(\ref{basicmetric}) because, as far as ${\cal K}$ depends
on time, this metric does not represent a spatially homogeneous
space-time, in the sense that the components of the four dimensional
curvature tensor in a local frame are functions of $t$ and $\xi$. The
existence of the non-vanishing component of the Ricci tensor, $R_{t
\xi} = [\xi \dot{k}(t)]/[a(t)N(t)]\not=0$, is a consequence of this
fact. Hence, we were forced to introduce a {\em midisuperspace} model
having a non-vanishing shift function $N_{\xi}(\xi, t)$. The metric
was written as
\begin{eqnarray}
\label{gbasicmetric}
{\rm d}s^2=\biggl[-N^2(\xi,t)+ \frac{N_{\xi}^2(\xi,t)}{a^2 (\xi,
t)}\biggr] {\rm d}t^2+2N_{\xi}(\xi,t){\rm d}\xi {\rm d}t \nonumber \\
+a^2(\xi,t)\biggl[{\rm d}\xi ^2+ {\sigma}^2(\xi ,t) {\rm d}\Omega
^2(\theta ,\varphi )\biggl].  
\end{eqnarray}
The first step was to carry out the quantization of this {\em
midisuperspace} model. Then, in a second step, we took into account in
the quantum solutions the restrictions on the variables $a$ and
$\sigma$ which, from Eq.~(\ref{gbasicmetric}), allows us to recover
the metric~(\ref{basicmetric}). Consistency requires that we first
treat the {\em midisuperspace} problem in order to come back to the
{\em minisuperspace} model afterward. Correspondingly, the Wheeler-De
Witt equation of the model remains a functional differential equation
rendering the findings of general exact solutions problematic. In
Ref.~\cite{LMPS}, only two particular semi-classical solutions were
found for which the possibility of a topology change at the quantum
level was explicitly demonstrated.

\par 

The present article aims at investigating the quantum behavior of the
system described by the metric~(\ref{basicmetric}) by generalizing the
{\em midisuperspace} description to a full superspace description in
order to get more general conclusions than those reached in
Ref.~\cite{LMPS}.

\par

Quantities of great interest for this purpose are the Green functions
since they completely characterize the quantum evolution of the
system. For a point particle, the Green function $G(x_{\rm f},t_{\rm
f};x_{\rm i},t_{\rm i})$ represents the probability amplitude to find
the particle at point $x_{\rm f}$ and time $t_{\rm f}$, knowing that
it was at point $x_{\rm i}$ at the initial time $t_{\rm i}$. In fact,
it is necessary to smooth out the Green function because, defined as
before, it is not square integrable. Physically, this is due to the
fact that the particle can never be localized exactly.  In the present
context, a Green function can also be defined and a similar
interpretation can be made. In that case, the wave function depends on
the volume of the spacelike sections $y \propto a^3$, on the three
curvature $R$, and on a dust field $\chi (t)$ (rigorous definitions
are given below), which plays the role of time: $\Psi =\Psi (y,R,\chi
)$, see Ref.~\cite{KT}.  The Green function $G(y_{\rm f},R_{\rm f},
\chi _{\rm f};y_{\rm i},R_{\rm i},\chi _{\rm i})$ can now be defined
as the probability amplitude of having a volume $y_{\rm f}$ and a
three-curvature $R_{\rm f}$ at ``time'' $\chi _{\rm f}$ knowing that,
at initial $\chi _{\rm i}$, the space volume was $y_{\rm i}$ and the
three-curvature $R_{\rm i}$. Therefore, the Green function defined
above represents the amplitude of probability for a topology
transition as soon as $R_{\rm f}$ has not the same sign as $R_{\rm
i}$. This is the quantity that we would like to calculate.

\par 

However, a full calculation of the Green functions is probably beyond
our present computational capability and therefore, we are forced to
rely on some approximation methods. Firstly, we will restrict
ourselves to a semi-classical approximation of the Wheeler-De Witt
equation. In this framework, the wave function will be given by the
semi-classical wave function $\Psi=\exp (iS/\hbar)$. In the case of a
topology change, the phase $S$ has a non-vanishing imaginary part, as
in ordinary quantum tunneling, because topology change cannot be
obtained classically, see Ref.~\cite{Ge}. Secondly, we will assume
that the spatial sections of our Universe model are made of compact
homogeneous backgrounds together with small perturbations around
them. As we are interested in large scale quantum changes of topology
in the homogeneous background and not on the small scale ones, typical
of Wheeler's space-time foam, we will use the long-wavelength
approximation, developed in Refs.~\cite{sal1,sal2,sal3,sal4}, which is
valid whenever the characteristic scale of the problem is much bigger
then the Hubble radius, and permits a complete analysis of the
semi-classical wave function and Green functions in the full
superspace. This will allows us to exhibit explicit analytical
expressions for the Green functions. In particular, we will
demonstrate that, for some transitions, the corresponding Green 
function vanishes, establishing thus selection rules for topology
changes.

\par 

This article is organized as follows. In the second section, we
briefly review the long wavelength approximation and the evaluation of
the Hamilton-Jacobi function up to the second order (first order in
the curvature). Then, in the third section, we perform the
calculations in the case of real metrics and show that there is no
quantum change of topology in this case. We also present a
generalization of the results obtained in Ref.~\cite{sal3}. In the
fourth section, we argue that complex metrics are necessary in order
to obtain quantum changes of topology. We explicitly calculate the
corresponding Green functions and discuss the main consequences that
can be drawn from their expressions. In the last and fifth section we
present our conclusions.

\section{The long wavelength approximation for quantum gravity 
with dust}
\paragraph*{}
In this section, we discuss the long-wavelength approximation
technique used to solve the Hamiltonian-Jacobi equation which is
nothing but the equation for the phase functional in a
Wentzel-Kramer-Brillouin (WKB) semi-classical approximation to
canonical quantum gravity. This technique consists in the expansion of
the phase functional (or the generating functional in a Hamiltonian
formulation of General Relativity) in a series of spatial gradients
and was introduced by Salopek, Stewart and Parry~\cite{sal1,sal2,sal3}
as a method for solving the full Hamilton-Jacobi equation for gravity
interacting with matter. In Ref.~\cite{sal3} these authors were able
to exhibit a recurrence formula from which they could obtain solutions
to the Hamilton-Jacobi equation to any order of approximation for
matter made of scalar and dust fields. This will be crucial for our
treatment because we shall use the second order solution to obtain the
Green functions associated with the WKB wave functional. As we will
show, the solution in the second order approximation will allow us to
describe changes of topology because the terms in the second order
expansion of the WKB phase functional explicitly contain spatial
curvature terms. As already mentioned, changes in the sign of these
terms indicate a change of topology.

\par

Let us now briefly describe the formalism introduced in
Refs.~\cite{sal1,sal2,sal3}. The action for General Relativity
interacting with a dust field $\chi$ is
\begin{eqnarray}
\label{action}
S &=& \int {\rm d}^4x \ \sqrt{-g} 
\biggl[\frac{1}{2\kappa }{}^{(4)}R-\frac{n}{2m}
\biggl(g^{\mu\nu}\partial_{\mu}\chi\partial_{\nu}\chi 
+m^2\biggr) 
\nonumber \\
& & - V(\chi )\biggr]\, ,
\end{eqnarray}
where $\kappa \equiv 8\pi /m_{\rm Pl}^2$, $m_{\rm Pl}$ being the
Planck mass. $^{(4)}R$ is the Ricci scalar of the space-time metric
$g_{\mu\nu}$, and $\chi$ is a velocity potential for irrotational dust
particles with rest mass $m$. $V(\chi )$ is a potential for the dust
field $\chi$ and $n$ is the rest number density of the dust
particles. The dust field $\chi$ defines the four velocity field for
dust particles by
\begin{equation}
\label{4vel}
u^{\mu } = -g^{\mu\nu} \frac{1}{m}\chi_{,\nu}\, 
\end{equation}
so that $\chi =$ const. will determine a congruence of spacelike
hypersurfaces foliating the space-time (notice that
$g^{\mu\nu}\chi_{,\mu}\chi_{,\nu} = -m^2)$. Therefore, the dust field
can be used as the time variable for our model, as well as in the
Schr\"odinger-type equation obtained from the Wheeler-De Witt equation.
\par In the Arnowitt-Deser-Misner formalism the line element reads
\begin{equation}
{\rm d}s^2 = (-N^2+\gamma^{ij}N_iN_j) {\rm d}t^2+2N_i \  
{\rm d}t {\rm d}x^i 
+\gamma_{ij}{\rm d}x^i {\rm d}x^i \; ,
\end{equation}
where $N$ and $N_i$ are the lapse and shift functions, respectively,
and $\gamma_{ij}$ is the three-metric of the spacelike
hypersurface. Then, the action~(\ref{action}) can be rewritten as
\begin{equation}
S = \int {\rm d}^4x ({\pi}^{\chi } \dot{\chi} + \pi^{ij}
\dot{\gamma}_{ij} - N{\cal H} - N^i {\cal H}_i) \; ,
\end{equation}
where $\pi^{ij}$ are the momenta conjugate to $\gamma_{ij}$ and
$\pi^{\chi}= n\gamma^{1/2}(1+\chi_{,i} \chi^{,i}/m^2)^{1/2}$ is the
momentum conjugate to the dust field $\chi$. The quantities ${\cal
H}$, ${\cal H}_i$ are given by the following expressions
\begin{eqnarray}
\label{superH}
{\cal H} &=&  {\kappa } \ \gamma^{-1/2} \pi^{ij} \pi^{k\ell}
(2\gamma_{jk}\gamma_{\ell i} - \gamma_{ij}\gamma_{k\ell}) 
\nonumber \\
&-&\frac{1}{2\kappa } \gamma^{1/2} R + (m^2+\chi_{,i}\chi^{,i})^{1/2}
\pi^{\chi}+\gamma^{1/2} V(\chi ) \; , \\
\label{supercons}
{\cal H}_i &=& -2 (\gamma_{i\ell}\pi^{\ell k})_{,k} + \pi^{\ell
k}\gamma_{\ell k ,i} + \pi^{\chi } \chi_{,i} \; ,
\end{eqnarray}
where $R$ is the Ricci scalar of the three-metric $\gamma _{ij}$.

\par

At this point it is well worth making some dimensional analysis, in
order to connect the conditions of validity of the long-wavelength
approximation to the Hubble radius of the Universe. From
Eq.~(\ref{superH}), we can see that the long-wavelength approximation
implies that the first term on the right-hand side is much larger than
the second term, at first order in the approximation. From the
expression for $\pi^{ij}=(K^{ij}-h^{ij}K)/\kappa$, see
Ref.~\cite{LMPS}, we can see that $\pi^{ij}$ has the dimension
$L^{-3}$ (in the system of units where $\hbar=c=1$) where $L$ stands
for a length. More explicitly,
\begin{eqnarray}
\nonumber
\pi^{ij}\sim \frac{1}{\ell_{\rm Pl}^{2} \ell _{\rm H}} ,
\end{eqnarray}
where $\ell_{\rm Pl}=1/m_{\rm Pl}$ is the Planck length and $\ell
_{\rm H}$ is essentially the Hubble radius. Therefore, the validity of
the long-wavelength approximation imposes that, see Eq.~(\ref{superH}),
\begin{eqnarray}
\nonumber
\lambda \gg \ell _{\rm H},
\end{eqnarray}
where the wavelength $\lambda$ is a typical length associated with the
scales where spatial gradients are relevant. In other words, $\lambda
$ must be greater than the characteristic Hubble radius of the
model. In our scheme, topology change is a large-scale (global)
phenomenon and the long-wavelength approximation is therefore the
appropriate approximation to probe it. The above condition, obviously,
should not contain the Planck length because Eq.~(\ref{superH}) is a
classical equation. Furthermore, in principle, there is no problem in
applying this condition to closed models, even when their volumes
becomes close to the Planck volume, (where, anyway, our semi-classical
approximation should not be valid) if the Hubble radius at this epoch
is smaller than the Planck length.

\par 

At the quantum level the above system may be quantized following the
Dirac's prescription~\cite{Dir}. The super-Hamiltonian and
super-momentum constraints given by Eqs.~(\ref{superH}) and
(\ref{supercons}) become operators and, when applied to the wave
functional of the system, result in two relations which express that
only a restricted region of the Hilbert space of wave functionals
contains the physical states of the theory:
\begin{equation} 
{\cal H} \Psi = 0 \, ,\quad {\cal H}_i \Psi = 0 \; .
\end{equation} 
The first equation is the well-known Wheeler-De Witt equation whereas
the second one is the so-called quantum momentum constraint. In what
follows, our treatment of quantum changes of topology will be done at
the level of the semi-classical approximation. This means that the
wave functionals are assumed to be of the WKB form, namely
\begin{equation}
\label{wkb}
\Psi = {\rm e}^{{iS}/{\hbar}},   
\end{equation}
where $S$ is the action. At order $\hbar^0$, Eqs.~(\ref{superH}) and 
(\ref{supercons}) reduce to the Hamilton-Jacobi functional equation
\begin{widetext} 
\begin{equation}
\label{hj}
{\cal H}(x)=\gamma^{-1/2} \kappa \ 
\frac{\delta S}{\delta\gamma_{ij}(x)}\
\frac{\delta S}{\delta\gamma_{k\ell}(x)}
[2\gamma_{i\ell}(x)\gamma_{jk}(x)-\gamma_{ij}(x)\gamma_{k\ell}(x)]
+ \sqrt{m^2+\gamma^{ij}\chi_{,i}\chi_{,j}}\ \frac{\delta S}{\delta\chi
(x)} + \gamma^{1/2} \ V(\chi ) - \frac{1}{2\kappa} \ 
\gamma^{1/2}R = 0 \; ,
\end{equation}
\end{widetext}
and to the momentum constraint equation
\begin{equation}
\label{nense30}
{\cal H}_i(x)=-2 \left[\gamma_{ik} \ \frac{\delta S}{\delta
\gamma_{kj}(x)}\right]_{,j} + \frac{\delta S}{\delta\gamma_{\ell
k}(x)} \ \gamma_{\ell k,i} + \frac{\delta S}{\delta \chi} \ 
\chi_{,j}
= 0 \; ,
\end{equation}
the solution of which is the phase $S$ of the WKB wave functional, see
Eq.~(\ref{wkb}). In Ref.~\cite{sal3}, Parry, Salopek and Stewart have
been able to derive solutions for these equations using the long
wavelength approximation. The method is based on the so-called spatial
gradient expansion of the phase functional $S$. It consists in
expanding $S$ in a series of terms according to the number of spatial
gradients they contain, namely $S = S^{(0)} + S^{(2)} + S^{(4)} +
\cdots$. As a consequence, the Hamilton-Jacobi equation can be grouped
in terms with an even number of spatial derivatives ${\cal H} = {\cal
H}^{(0)} + {\cal H}^{(2)} + {\cal H}^{(4)} + \cdots$. At zeroth order,
the Hamilton-Jacobi equation reduces to
\begin{eqnarray}
\label{hj0}
{\cal H}^{(0)} &=& \gamma^{-1/2} \kappa \ \frac{\delta
S^{(0)}}{\delta\gamma_{ij}} \ \frac{\delta
S^{(0)}}{\delta\gamma_{k\ell}} (2\gamma_{jk}\gamma_{\ell
i}-\gamma_{ij}\gamma_{k\ell}) \nonumber \\
& & +m \frac{\delta
S^{(0)}}{\delta \chi} +\gamma^{1/2}V(\chi )=0\, . 
\end{eqnarray}
For a dust field $\chi$ with a vanishing potential, $V(\chi )=0$, the
invariance under diffeomorphisms of the generating functional suggests
a solution of the form
\begin{equation}
\label{s0}
S^{(0)} = -\frac{2}{\kappa} \int {\rm d}^3x \gamma^{1/2} H(\chi ) \; .
\end{equation}
This functional satisfies the Hamilton-Jacobi relation if $H(\chi)$ is
a solution of $H^2=-(2m/3)\partial H/\partial \chi$ yielding
\begin{equation}
\label{huble}
H(\chi ) = \frac{2m}{3(\chi -\tilde{\chi})}\, ,
\end{equation}
where $\tilde{\chi}$ is an integration constant.  The functional
$S^{(0)}$ is clearly invariant under coordinate transformations and,
therefore, satisfies Eq.~(\ref{nense30}). In addition, it contains no
spatial derivatives. For the class of generating functionals given by
Eq.~(\ref{s0}), the second-order Hamilton-Jacobi equation reads
\begin{eqnarray}
\label{h2}
{\cal H}^{(2)} &=& 2H \gamma_{ij} \ \frac{\delta
S^{(2)}}{\delta\gamma_{ij}} + m\frac{\delta S^{(2)}}{\delta\chi} +
\frac{1}{2m} \gamma^{ij} \chi_{,i}\chi_{,j} \ \frac{\delta
S^{(0)}}{\delta\chi} 
\nonumber \\
& & - \frac{1}{2\kappa} \gamma^{1/2} R = 0 \; .
\end{eqnarray}
In Ref. \cite{sal3}, Parry, Salopek and Stewart obtained a
diffeomorphisms invariant solution for $S^{(2)}$ which can be written
as
\begin{equation}
\label{s1}
S^{(2)}=\frac{1}{\kappa}\int{\rm d}^3x \gamma^{1/2} J(\chi )R \; ,
\end{equation}
where the function $J(\chi )$ is obtained by substituting
Eq.~(\ref{s1}) into Eq.~(\ref{h2}) and using the zeroth order solution
(\ref{s0}). This leads to
\begin{equation}
\label{Jeq}
HJ+m\frac{\partial J}{\partial \chi}=\frac{1}{2},
\end{equation}
which yields for $S^{(2)}$
\begin{equation}
\label{s2}
S^{(2)} = \frac{1}{\kappa} \int {\rm d}^3x \ 
\gamma^{1/2} \biggl[\frac{3}{10 m}
(\chi-\tilde{\chi}) + D(\chi - \tilde{\chi})^{-2/3}\biggr] R \; ,
\end{equation}
where $D$ is an arbitrary integration constant. We notice the fact
that $S^{(2)}$ contains the three-curvature term $R$ whereas this is
not the case for $S^{(0)}$. Since a topology change is signaled by a
change of sign of the three-curvature $R$, it is necessary to push the
long wavelength expansion to the second order in order to have non
trivial effects. In other words, the quantum topology changes can only
be studied at this (second) order.

\section{The Green function with Lorentzian metrics}

\subsection{General Derivation}

Let us now consider the semi-classical wave function up to the second
order in the long wavelength approximation $\Psi =\exp
\{i[S^{(0)}+S^{(2)}]/\hbar \}$ with $S^{(0)}$ and $S^{(2)}$ given by
Eqs.~(\ref{s0}) and (\ref{s2}), as explained in the previous
section. We restrict ourselves to geometries of the form given in
Eq.~(\ref{basicmetric}). Then, the integrals in Eqs.~(\ref{s0}) and
(\ref{s2}) can be performed exactly yielding
\begin{eqnarray}
\label{wave2}
\Psi(y,R,\chi) &=& \exp \biggl\{\frac{i}{\hbar \kappa}\biggl[-2y H +
\frac{3yR(\chi -\tilde{\chi})}{10m} \nonumber \\
& & +\frac{yRD}{(\chi -\tilde{\chi})^{2/3}}\biggr]\biggr\},
\end{eqnarray}
where the quantity $y$ given by
\begin{equation}
\label{ydef}
y\equiv\int{\rm d}^3x\gamma ^{1/2} = V a^3
\end{equation}
is the total volume associated with the metric~(\ref{basicmetric}).
$V$ is the constant volume of the cell corresponding to the
homogeneous polyhedron of the tessellation of the spatial sections that
are compactified. As previously, $H$ can be expressed as $H=2m/[3(\chi
-\tilde{\chi})]$ and $R$ is the three-curvature which can be written
as $R=6{\cal K}/a^2$. The Green functions $G$ are defined by
\begin{eqnarray}
\label{green2}
\Psi(y,R,\chi) &\equiv &\int^{\infty}_{0} {\rm d}y'
\int^{\infty}_{-\infty}{\rm d}R'G(y,R,\chi;y',R',\chi ')
\nonumber \\
& & \times \Psi(y',R',\chi ').
\end{eqnarray}
Our goal is now to establish the expression for the Green function
itself from the above equation. For this purpose, it is convenient to
use a new variable $v$, which is a combination of $y$ and $R$, namely
$v\equiv yR$. To obtain the Green function, we use the fact that
Eq.~(\ref{green2}) is valid for any value of the integration constants
$\tilde{\chi}$ and $D$. However, instead of the integration constants
$\tilde{\chi}$ and $D$, it will be more appropriate to work with new
integration constants defined by the following relations (for fixed
$\chi '$)
\begin{eqnarray}
H'(\tilde{\chi })& \equiv & \frac{2m}{3(\chi ' -\tilde{\chi})} 
\\
J '(\tilde{\chi },D) &\equiv & 
\frac{3}{10m}(\chi ' -\tilde{\chi})+
D(\chi ' -\tilde{\chi})^{-2/3}\, .
\end{eqnarray}
We have just replaced the two constants $(\tilde{\chi },D)$ by two new
constants $(H',J ')$. Let us notice that $J '$ is in fact $J(\chi ')$,
see Eqs.~(\ref{s1}) and (\ref{s2}), hence its name. If we inverse the
above relations, one arrives at
\begin{eqnarray}
\label{invcst}
\tilde{\chi }(H') &=& \chi '-\frac{2m}{3H'}\, ,
\\ 
\label{invcst2}
D(H',J ') &=& J '\biggl(\frac{2m}{3H'}\biggr)^{2/3}
-\frac{3}{10m}\biggl(\frac{2m}{3H'}\biggr)^{5/3}\, .
\end{eqnarray}
Let us stress again that these expressions are defined for a fixed
value of $\chi '$. With these new definitions and the explicit
expression of the wave function given in Eq.~(\ref{wave2}),
Eq.~(\ref{green2}) now reads
\begin{widetext}
\begin{eqnarray}
\label{simple}
\exp \biggl(\frac{i}{\hbar \kappa}\biggl\{-2y H +
\frac{3v[\chi -\tilde{\chi}(H')]}{10m} +
\frac{vD(H',J ')}{[\chi -\tilde{\chi}(H')]^{2/3}}
\biggr\}\biggr)
&=& \int^{\infty}_{0} \frac{{\rm d}y'}{y'} 
\int^{\infty}_{-\infty}{\rm d}v'G\biggl(y,\frac{v}{y},
\chi;y',\frac{v'}{y'},\chi '\biggr) 
\nonumber \\
& &\times \exp \biggl[\frac{i}{\hbar \kappa}(-2y' H' + 
v'J ')\biggr] \, .
\\ \nonumber
\end{eqnarray}
\end{widetext}
We see the effect of having introduced the ``new'' integration
constants: the left-hand-side of the previous equation now depends on
both $\chi $ and $\chi '$. In order to extract
$G(y,v/y,\chi;y',v'/y',\chi ')$, we multiply the previous equation by
the factor
\begin{equation}
\exp \biggl[\frac{i}{\hbar \kappa}(2y'' H' - v''J ')\biggr],
\end{equation}
and integrate it over $J '$ and $H'$ from $-\infty$ to $\infty$. The
result of this integration on the right-hand-side of
Eq.~(\ref{simple}) is very simple, yielding two Dirac delta functions whose
arguments are $v'-v''$ and $y'-y''$. This allows us to perform the
integration over $v'$ and $y'$ immediately. The result reads
\begin{widetext}
\begin{eqnarray}
\label{main}
\nonumber 
&& G\biggl(y,\frac{v}{y},\chi;y'',\frac{v''}{y''},\chi '\biggr)\\
&=&\frac{y''}{2 \pi^{2} \hbar^{2} \kappa^{2}}\int^{\infty}_{-\infty} 
{\rm d}H' \int _{-\infty }^{\infty }{\rm d}J '
\exp \biggl\{\frac{i}{\hbar \kappa}\biggl[-\frac{2yb H'}{H'+b} 
+ 2y''H'- v''J ' +
v\biggl(J '-\frac{1}{5H'}\biggr)
\biggl(\frac{b}{H' + b}\biggr)^{2/3}+\frac{v}{5}\biggl(\frac{1}{b}
+\frac{1}{H'}\biggr)\biggr]\biggr\}\; ,
\nonumber \\
& & 
\end{eqnarray}
\end{widetext}
where $b\equiv 2m/[3(\chi - \chi ')]$. The first term in the argument
of the exponential has been obtained by using the relation
$H=2m/[3(\chi -\tilde{\chi })]$ and then Eqs.~(\ref{invcst}) and
(\ref{invcst2}). The two last terms have been obtained by using the
explicit expressions of $\tilde{\chi }$ and $D$, see
Eqs.~(\ref{invcst}) and (\ref{invcst2}). The above equation is the
main result of this section. Clearly, the central issue is now the
calculation of the integrals over $J'$ and $H'$. {}From
Eq.~(\ref{main}), we see that the integral over $H'$ is to be
performed along the real axis. However, in the next subsection,
dealing with the case of flat hypersurfaces, the integration over $H'$
will be performed in the complex plane, along a closed contour. It is
important to notice that this procedure is just a technical trick,
allowing us to simplify the calculation of the integral by the use of
the Cauchy theorem, but it has no deep physical meaning. On the
contrary, in section IV, the integration in the complex plane has a
physical meaning and, as we will argue, is in fact necessary. Of
course, all the following complex contours turn out to be, after
having taken the appropriate limits, equivalent to the integration
along the real axis as given in Eq.~(\ref{main}) .

\subsection{The Case of Flat Hypersurfaces}

In this section we treat the case where either the starting
hypersurface or the resulting hypersurface (after the topology change)
is flat. 

\par

Let us first start with the case $R=0$ and $R''\neq 0$, hence $v=0$
but $v''\neq 0$. In this case, the integration over $J '$ is very
simple and yields the Dirac function $\delta(v'')$. Consequently, in
order to obtain a non-vanishing Green function, one must take $v''=0$,
that is to say $R''=0$. This means that the three-curvature remains
the same (i.e. zero) and there is obviously no topology change. At
this point, however, we go further and depict the calculation of the
Green function despite the fact that there is no change of topology
and despite the fact that the final result is already known, see
Ref.~\cite{sal4}. The main reason for doing this is that this will
exemplify techniques that will be used in this paper later on. This
calculation is therefore a warm-up. Eq.~(\ref{main}) now reduces to
\begin{eqnarray}
\label{mainflat}
& & G(y,0,\chi;y'',0,\chi ')
\nonumber \\
&=& \frac{y''}{2 \pi^{2} \hbar^{2} \kappa^{2}}
\int^{\infty}_{-\infty} {\rm d}H' 
\exp \biggl[\frac{i}{\hbar \kappa}\biggl(\frac{2yb H'}{H'+b} 
+ 2y''H'\biggr)\biggr]\, . 
\nonumber \\
& & 
\end{eqnarray}
As mentioned before, the integral in the 
\begin{figure}[t]
\begin{center}
\hspace{-1.0cm}
\includegraphics[width=8.5cm]{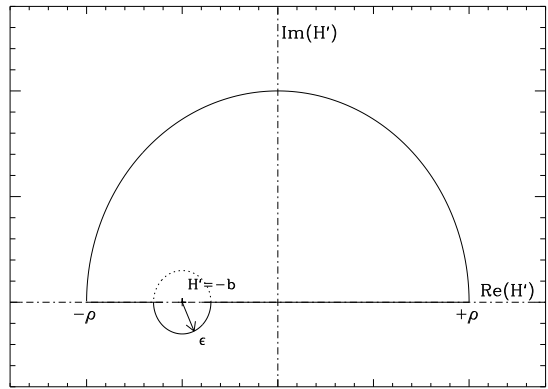}
\caption{The contour of integration ${\cal C}$ in the complex plane
$[\Re(H'),\Im(H')]$. The contour chosen in this article is the solid
line contour. Because of the Cauchy's theorem, ${\cal C}$ is
equivalent to the small circle contour centered at $H'=-b$.  }
\label{contour1}
\end{center}
\end{figure}
right-hand-side of Eq.~(\ref{mainflat}) can be calculated by extending
$H'$ to the complex plane with the choice of the integration contour
represented by the solid line in Fig.~\ref{contour1}.  This contour
contains the pole $H'=-b$ of the exponential function appearing in the
right-hand-side of Eq.~(\ref{mainflat}). The fact that $H'=-b$, which
is a pole of the argument of the exponential, is also a pole of the
exponential itself can easily be seen if one uses the Taylor expansion
of the exponential function. When the limits $\rho \rightarrow +\infty
$ and $\epsilon \rightarrow 0$ are taken, see Fig.~\ref{contour1}, the
integral over $H'$ in the right-hand-side of Eq.~(\ref{mainflat})
reduces to an integration on the real axis, the contribution coming
from the big semicircle in the upper-half complex plane going to zero
as a consequence of the Jordan's lemma. Because of the Cauchy's
theorem, the integration along ${\cal C}$ is in fact equivalent to the
integration over a small circle centered around $H'=-b$, see
Fig.~\ref{contour1}. This was the contour considered in
Ref.~\cite{sal4} which is therefore equivalent to the contour ${\cal
C}$ used in this article. As a consequence, one obtains
\begin{eqnarray}
\label{simple2}
& & G(y,0,\chi;y'',0,\chi ') =\frac{y''}{2\pi ^2 \hbar^2 \kappa ^2}
\nonumber \\
& \times &  \int _{\cal C} {\rm d}H'
\exp \biggl[\frac{i}{\hbar \kappa}\biggl(-\frac{2yb H'}{H'+b} 
+ 2y''H'\biggr)\biggr]\, . 
\end{eqnarray}
It has been shown in Ref.~\cite{sal4} that the integration over $H'$
along the contour ${\cal C}$ can then be done explicitly. The exact
result is expressed in Eq.~(4.25) of Ref.~\cite{sal4} in terms of a
Bessel function of order one.  We do not present this result here
since this aspect of the calculation has been given in details in
Ref.~\cite{sal4}.

\par

Let us now consider the other possibility where $R\neq 0$, hence
$v\neq 0$ but now $R''=0$ ($v''=0$). Despite the fact that the
argument of the exponential function is now more complicated, the
integration over $J '$ can still be performed very easily and yields
the Dirac delta function
\begin{equation}
\label{delta0}
\delta \biggl[v\biggl(\frac{b}{H'+b}\biggr)^{2/3}\biggr]\, .
\end{equation}
Since the argument of the Dirac function has no zeros in $H'$, the
integration over $H'$ leads to a vanishing Green function. 

\par

Hence, we have established the following result: there is no quantum
change of topology from or to a flat spacelike hypersurface in the
homogeneous background, at least at second order in the
long-wavelength approximation. 

\subsection{The Case of Curved Hypersurfaces}

We now assume that both $R$ and $R'$, and hence $v$ and $v''$, do not
vanish. This time the integration over $J '$, in the right-hand-side
of Eq.~(\ref{main}), gives the following Dirac function [to be
compared with Eq.~(\ref{delta0})]
\begin{equation}
\label{delta}
\delta\biggl[v\biggl(\frac{b}{H'+b}\biggr)^{2/3}-v''\biggr]\; .
\end{equation}
Since $H'$ is real, the argument of the Dirac function has zeros if
and only if $R$ and $R'$ (or, equivalently, $v$ and $v''$) have the
same sign and, in this case, the zeros are given by
\begin{equation}
H'=H'{}_0^{\mp }\equiv -b
\biggl[1\mp \biggl(\frac{v}{v''}\biggr)^{3/2}\biggr]\, .
\end{equation}
If $R$ and $R'$ have not the same sign, the corresponding Green
function is zero and quantum topology change is forbidden. Therefore,
as in the flat case, there is no quantum change of topology.  

\par

Despite this fact, and for the sake of completeness, we now present
the resulting Green function since this represents a non trivial
generalization of the results obtained in Ref.~\cite{sal4}.  Let us
notice that this case is different from the one given in
Eq.~(\ref{mainflat}). There, we had a Dirac function of $v''$ only
and, therefore, it was possible to put the Dirac function outside the
integral over $H'$. Here, we face a different situation since, this
time, the Dirac function does depend on $H'$ (the second flat case ,
i.e. $v\neq 0$, also presented this feature but since the argument of
the Dirac function had no zeros, this case was in fact trivial). In
order to calculate the integral~(\ref{main}) for the present case, we
express the Dirac function as
\begin{eqnarray}
\label{deltamagic}
& & \delta\biggl[v\biggl(\frac{b}{H'+b}\biggr)^{2/3}-v''\biggr]
=\frac{3b}{2\vert v\vert}\biggl(\frac{v''}{v}\biggr)^{-5/2}
\nonumber \\
& & \times \biggl[\delta (H'-H'{}_0^{-})
+\delta (H'-H'{}_0^{+})\biggr]\, ,
\end{eqnarray}
and the integration becomes trivial. As a consequence, the Green
function is given by the sum of two contributions coming from the two
Dirac functions in Eq.~(\ref{deltamagic})
\begin{equation}
G_r = (G_+ + G_-)\Theta\biggl(\frac{{\cal K}}{{\cal K}''}\biggr) \; ,
\end{equation}
where $\Theta$ is the step function and where $G_{\pm}$ are given by
\begin{widetext}
\begin{eqnarray}
G_+ &=& \frac{3 m y''x^{5/2}}{\pi \hbar \kappa \mid v 
\mid (\chi-{\chi}'')}
\exp \biggl\{\frac{i}{\hbar \kappa}
\biggl[-\frac{2yb(x^{3/2}-1)}{x^{3/2}}
+2y''b(x^{3/2}-1)+\frac{{\cal K}y^{1/3}(x^{5/2}-1)}{5b(x^{3/2}-1)x}
\biggr]\biggr\}\; ,
\\
G_- &=& \frac{3 m y''x^{5/2}}{\pi \hbar \kappa \mid v 
\mid (\chi-{\chi}'')}
\exp \biggl\{\frac{i}{\hbar \kappa}
\biggl[-\frac{2yb(x^{3/2}+1)}{x^{3/2}}
-2y''b(x^{3/2}+1)+\frac{{\cal K} y^{1/3}(x^{5/2}+1)}{5b(x^{3/2}+1)x}
\biggr]\biggr\}\; ,
\end{eqnarray}
\end{widetext}
with $x\equiv(y/y'')^{1/3}$. This completes our study of the Green
functions in the case of real metrics. We have demonstrated that there
is no quantum change of topology in that case, as long as this
transition involves a change of sign in the curvature of the spatial
sections. Our formulation does not allow to signal a change of
topology when the transition involves compactified spacelike sections
with distinct topology but having the same sign of curvature, a fact
that would require a completely distinct and much more elaborated
approach to the problem. These conclusions are valid up to second
order in the gradient expansion.

\section{The Green function with complex metrics}

We have just seen in the previous section that there is no change of
topology if $H'$ remains real. This is linked to the fact that the
equation $v''=v[b/(H'+b)]^{2/3}$ has no solution on the real axis of $H'$ if
$v$ and $v''$ have opposite signs. It is worth recalling again that
the extension of $H'$ to the complex plane made in section III, in the
case of flat hypersurfaces, was a mere mathematical trick to calculate
the integrals, albeit all the poles inside the contours were located
on the real axis. On the other hand, the above equation possesses a
solution in the complex plane, namely
\begin{equation}
\label{complexroot}
H'=H'_0\equiv -b\biggl(1+
i\biggl\vert \frac{v}{v''}\biggr \vert ^{3/2}\biggr)\, 
\end{equation}
and, therefore, this strongly suggests to consider $H'$ as a true
complex variable. As we will discuss in the present section, the
complexification of $H'$, with its natural consequence namely a
complexification of the metric itself, is a physical requirement. This
will lead us to interpret topology change as a quantum tunneling
effect. In the case of complex metrics, the corresponding Green 
function needs to be modified and its expression can be obtained from
the expressions derived before. We now describe how this can be done.

\subsection{General derivation}

Changes of topology described by the metric~(\ref{basicmetric})
requires a complicated {\em midisuperspace} formulation of the
problem, as described in the Introduction and explained in more
details in Ref.~\cite{LMPS}. However, in the case of curved spatial
hypersurfaces, the {\em midisuperspace} formulation can be
circumvented and the problem greatly simplified by the introduction of
complex metrics. Let us now define the metric according to
\begin{equation}
\label{a1}
{\rm d}s^2=-N_{\mathrm c}^2(t){\rm d}t ^2+a^2(t)\biggl[{\rm d}z ^2+ \sin
^2z({\rm d}\theta ^2 + \sin ^2\theta{\rm d}\varphi ^2)\biggr]\; ,
\end{equation}
and allow $N_{\mathrm c}$ and $z$ to be complex.  Setting $N_{\mathrm
c}=N$ and $z=\xi$ to be real yields, let us say for the final
configuration, a positively curved spatial hypersurface in a
Lorentzian four-geometry, while requiring initially that $N_{\mathrm
c}=-iN$ and $z=-i\xi$ are purely imaginary yields a negatively curved
spatial hypersurface, also in a Lorentzian four-geometry [of course,
one can interchange the transition by choosing $\sinh$ instead of
$\sin$ in Eq.~(\ref{a1})]. Therefore, the transition from positively
to negatively curved spatial sections may be obtained with a mere
complexification of the metric, without changing the functional form
of the metric in Eq.~(\ref{a1}) or without needing to introduce a time
dependence on ${\cal K}$, as in Eq.~(\ref{basicmetric}). In other
words, the transition described above can be actually realized by
following trajectories in the space of complex metrics. Notice also
that, with the complexification procedure, there is an overall change
of sign in the $4$-metric while the transition from the initial to the
final configuration occurs , implying a passage through degenerate
metrics. This is typical of transitions with a change of topology as
pointed out in Ref.~\cite{hawking}. Let us also emphasize that the use
of complex trajectories in order to describe the tunneling effect is
standard and has been used elsewhere, for instance in the study of
chaotic potentials or in the calculation of the rotational spectra of
molecules, see Ref.~\cite{complexorb}.

\par

The change in the lapse function, $N_{\mathrm c}=-iN$, is equivalent
to a Wick rotation to the imaginary time $\tau=-it$, if one
understands $t$ as the proper time of observers whose trajectories are
orthogonal to the spacelike hypersurfaces. In this context, one needs
to adapt the ADM formalism presented at the beginning of this article
in order to obtain the Green function that will describe a topology
change. We now turn to this question.

\par

The Euclidean action defined as $I \equiv -iS[g_{\mathrm E}]$, see
Ref.~\cite{hawking}, is given by
\begin{eqnarray}
\label{Eaction}
I &=& -\int {\rm d}^4x \ \sqrt{g_{\mathrm E}} 
\biggl[\frac{1}{2\kappa }{}^{(4)}
R(g_{\mathrm E})
\nonumber \\
& & -\frac{n}{2m}
\biggl(g^{\mu\nu}_{\mathrm E}\partial_{\mu}\chi\partial_{\nu}\chi 
+ m^2\biggr)- V(\chi )\biggr] \, ,
\end{eqnarray}
where $g_{\mathrm E}$ denotes the Euclidean $4$-geometry whose line
element can be expressed as
\begin{equation}
{\rm d}s^2 = (N^2+\gamma^{ij}N_iN_j) {\rm d}t^2+2N_i \  {\rm d}t {\rm d}x^i
+\gamma_{ij}{\rm d}x^i {\rm d}x^i \; .
\end{equation}
Note that we have used the relation $\sqrt{-g_{\mathrm
c}}=-i\sqrt{g_{\mathrm E}}$, where $g_{\mathrm c}$ is a Lorentzian
metric complexified as in Eq.~(\ref{a1}). Having determined what the
action is, we can study the Hamiltonian formalism. Calculations very
similar to those performed in the standard ADM formalism show that the
Hamiltonian obtained from the Euclidean action given above reads
\begin{equation}
H_{\mathrm E} = N{\cal H}_{\mathrm E} + N^i {\cal H}_i \; ,
\end{equation}
where ${\cal H}_{\mathrm E}$ and ${\cal H}_i$ are respectively given
by
\begin{eqnarray}
\label{superHE}
{\cal H}_{\mathrm E} &=&  -{\kappa } \ \gamma^{-1/2} \pi^{ij} \pi^{k\ell}
(2\gamma_{jk}\gamma_{\ell i} - \gamma_{ij}\gamma_{k\ell}) -
\frac{1}{2\kappa } \gamma^{1/2} R 
\nonumber \\
& & + i(m^2+\chi_{,j}\chi^{,j})^{1/2}
\pi^{\chi}+\gamma^{1/2} V(\chi ) \; , \\
\label{superconsE}
{\cal H}_i &=& -2 (\gamma_{i\ell}\pi^{\ell k})_{,k} + \pi^{\ell
k}\gamma_{\ell k ,i} + \pi^{\chi } \chi_{,i} \; ,
\end{eqnarray}
and where, as before, $\pi^{ij}$ are the conjugate momenta to
$\gamma_{ij}$ and $\pi^{\chi}$ is the conjugate momentum to the dust
field $\chi$. The quantity $R$ is the Ricci scalar of the three-metric
$\gamma _{ij}$. From the expression of ${\cal H}_{\mathrm E}$ derived
above, one can establish the Euclidean Hamilton-Jacobi equation. It
reads
\begin{widetext}
\begin{eqnarray}
\label{hjE}
-\gamma^{-1/2} \kappa \ \frac{\delta S}{\delta\gamma_{ij}(x)}\
\frac{\delta S}{\delta\gamma_{k\ell}(x)}
[2\gamma_{i\ell}(x)\gamma_{jk}(x)-\gamma_{ij}(x)\gamma_{k\ell}(x)]
+i \sqrt{m^2+\gamma^{ij}\chi_{,i}\chi_{,j}}\ \frac{\delta S}{\delta\chi
(x)} + \gamma^{1/2} \ V(\chi ) - \frac{1}{2\kappa} \ \gamma^{1/2}R = 0 \; ,
\end{eqnarray}
\end{widetext}
while the momentum constraint equation remains the same as in
Eq.~(\ref{nense30}). Comparing Eq.~(\ref{hjE}) with Eq.~(\ref{hj}),
one can see that going to the imaginary time (by means of the Wick
rotation) changes the sign of the kinetic term in the gravitational
sector. This is a well-known modification when going from Lorentzian
to Euclidean signatures. On the other hand, the matter sector,
containing the term $\delta S/\delta\chi$, is multiplied by complex
number $i$. This latter change was also expected since $\chi$ plays
the role of time. Setting $\tau=-it$ implies $\chi_{\mathrm c}
=-i\chi$ after the Wick rotation, hence the above result.

\par

Having the Euclidean Hamilton-Jacobi equation and the momentum
equation at our disposal, we now seek solutions using the gradient
expansion. In the following, quantities with the index $\mathrm c$
refer to the metric $(\ref{a1})$ after the Wick rotation ($N_{\mathrm
c}=-iN$, $z$ complex) while quantities without index refer to the
(real) Lorentzian metric (\ref{a1}) ($N$=real, $\xi$ real). As it was
the case before, see Eq.~(\ref{s0}), the invariance by
reparametrization suggests a solution of the form
\begin{equation}
\label{s0E}
S^{(0)}_{\mathrm c} = -\frac{2}{\kappa} \int {\rm d}^3x 
\gamma_{\mathrm c}^{1/2} H_{\mathrm c}(\chi ) \, .
\end{equation}
Inserting this form into the Euclidean Hamilton-Jacobi equation
yields, at the zero order, $H_{\mathrm c}^2=(2im/3)\partial H_{\mathrm
c}/\partial \chi$. This equation should be compared with the equation
written after Eq.~(\ref{s0}). The difference between those two
equations is a factor $-i$ as expected. The corresponding solution can
be written as $H_{\mathrm c}=-iH$, where $H(\chi ) = 2m/[3(\chi
-\tilde{\chi})]$. Again, this solution bears some close resemblance
with the solution of Eq.~(\ref{huble}).

\par

At second order, the diffeomorphism invariant solution reads 
\begin{equation}
\label{s1E}
S^{(2)}_{\mathrm c}=\frac{1}{\kappa}\int{\rm d}^3x 
\gamma _{\mathrm c} ^{1/2} J_{\mathrm c}(\chi )R_{\mathrm c} \, ,
\end{equation}
and is similar to the solution of Eq.~(\ref{s2}). Applying the same
method as before, i.e. inserting the above equation into the Euclidean
Hamilton-Jacobi equation, yields the following relation
\begin{equation}
\label{JeqE}
-H_{\mathrm c}J_{\mathrm c}+im\frac{\partial J_{\mathrm c}}
{\partial \chi }=iHJ_{\mathrm c}+im\frac{\partial J_{\mathrm c}}
{\partial \chi }=\frac{1}{2}\, .
\end{equation}
Since we have already established the expression of $H_{\mathrm c}$,
the above relation is a differential equation for the quantity
$J_{\mathrm c}$ only. The solution is very similar to the one found in
section II of this paper. It reads $J_{\mathrm c}=-iJ$, where, as can
be deduced from Eq.~(\ref{s2}), one has
\begin{equation}
\label{s2E}
J (\tilde{\chi },D)= \frac{3}{10 m}
(\chi-\tilde{\chi}) + D(\chi - \tilde{\chi})^{-2/3}\; .
\end{equation}

\par

We now have at our disposal the Euclidean action at zeroth and second
order in full generality, i.e. for any metric tensor. As before, we
now restrict our considerations to the metrics of the
form~(\ref{a1}). This allows us to perform the integrals~(\ref{s0E})
and (\ref{s1E}) exactly. However, in the Euclidean case, one has to
pay attention to the following fact. When the transformation $z=-i\xi$
is made, the three-metric gets modified by an overall sign (besides
the change from $\sin$ to $\sinh$). This implies that $\gamma_{\mathrm
c}^{1/2} =-i\gamma ^{1/2}$ and, hence, the volume $y_{\mathrm c}$ can
be written as
\begin{equation}
\label{ydefE}
y_{\mathrm c}\equiv\int{\rm d}^3x\gamma_{\mathrm c}^{1/2} = -iy \; .
\end{equation}
For the same reason, the three-curvature satisfies $R_{\mathrm
c}\equiv R(\gamma^{\mu\nu}_{\mathrm c}) =-R(\gamma^{\mu\nu})\equiv
-R$. Combining the previous results, one has $v_{\mathrm c} =
y_{\mathrm c} R_{\mathrm c} = iyR = iv$.

\par

We are now in a position where we can calculate the semi-classical
wave function in the long wavelength approximation at the time
$\chi$. The result reads
\begin{eqnarray}
\label{psic}
\Psi(y_{\mathrm c}, R_{\mathrm c}, \chi )&=&
\exp \biggl[\frac{i}{\hbar \kappa}(-2y_{\mathrm c} H_{\mathrm c}
+v_{\mathrm c}J_{\mathrm c})\biggr]
\\ 
&=&\exp \biggl[\frac{i}{\hbar \kappa}(2yH + vJ )\biggr]\; .
\\
& & \nonumber
\end{eqnarray}
This solution should be compared with Eq.~(\ref{wave2}). One notices
that there is a difference, namely a change of sign in the first term
of the exponential. As we are going to show below, this simple change
of sign turns out to be crucial in order to obtain a Green function
which allows a quantum change of topology.

\par

The last step is to calculate the Green function explicitly. This
Green function describes a transition from a situation where the wave
function is expressed by the standard formula, Eq.~(\ref{wave2}), to a
case where the wave function is modified by the above-mentioned change
of sign, see Eq.~(\ref{psic}) (or vice versa if one wishes to consider
the opposite transition). As a consequence, in order to obtain the
Green function, we just need to repeat the calculations performed
between Eq.~(\ref{green2}) and Eq.~(\ref{simple2}), bearing in mind
that, in the final (or initial) configuration, the wave function is
now described by Eq.~(\ref{psic}). We obtain the following result
\begin{widetext}
\begin{eqnarray}
\label{A1}
G_i(y,R,\chi;{y}',{R}',{\chi}') = 
\frac{{y}'}{2\pi ^2 \hbar^2 \kappa ^2}
\int _{\cal C} {\rm d}{H}' \int^{\infty}_{-\infty}{\rm d}{J}'
\exp \biggl\{\frac{i}{\hbar \kappa}\biggl[-2y H - 2{y}'{H}'-v'J '
+\frac{3v(\chi  -{\tilde{\chi}})}{10m} +
\frac{vD}{(\chi -{\tilde{\chi}})^{2/3}}\biggr]\biggr\} \; .
\end{eqnarray}
\end{widetext}
At this point, one should compare the above result with
Eq.~(\ref{simple2}). Being given the previous considerations, it does
not come as a surprise that the sign of the second term in the
exponential is different. The origin of the difference is clearly the
modified wave function of Eq.~(\ref{psic}). Apart from this (crucial)
difference, the previous expression is identical to
Eq.~(\ref{main}). Let us also notice that, in order to derive the
Green function, we have already assumed that the contour ${\cal C}$
reduces to an integration along the real axis. This allows us to
extract the Green function directly. We will show below that this is
indeed the case.

\subsection{Explicit calculation of the Green function}

Equipped with the modified Green function~(\ref{A1}), we now return to
our main subject, {\it i.e} the study of the quantum topology
changes. We now consider the case where $v$ and $v'$ have opposite
signs, namely sign($v$)$=-$sign($v'$).

\par

\begin{figure}[t]
\begin{center}
\hspace{-1.0cm}
\includegraphics[width=8.5cm]{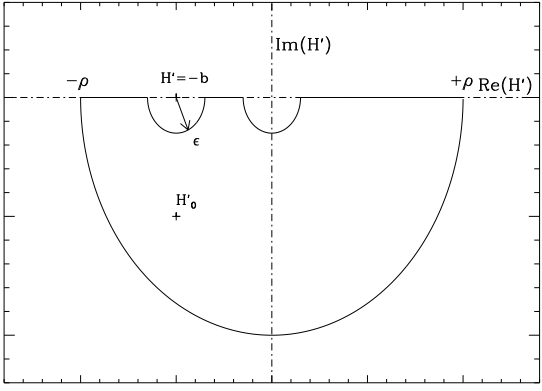}
\caption{The contour of integration ${\cal C}$ in the complex plane
$[\Re(H'),\Im(H')]$ in the case of curved hypersurfaces for the
modified Green function. The contour chosen is the solid line contour
which contains the complex pole of the Dirac function $H_0^{'}$ but
avoids the poles of the exponential function, namely $H'=0$ and
$H'=-b$.}
\label{contour3}
\end{center}
\end{figure}

The first step is to perform the integration over $J'$. This yields
the Dirac $\delta$-function of Eq.~(\ref{delta}) (except that one
should replace $v''$ by $v'$). This function has the root given in
Eq.~(\ref{complexroot}), i.e. $H'=H'_0$. The second step is the choice
of the contour ${\cal C}$. In the complex plane, integration of a
Dirac $\delta$-function can be represented by an integration along a
small circle around the roots of its argument. For this reason, the
integration over $H'$ will be performed along the contour which goes
from $-\rho $ to $\rho $ on the real axis, avoiding the poles of the
exponential function, $H'=-b$ and $H'=0$. The contour is closed by
considering a large semicircle in the half part of the complex plane
with negative imaginary part. This contour is represented in
Fig.~\ref{contour3}. It is equivalent to a small closed contour around
the pole $H'=H'_0$ because of the Cauchy theorem. In addition, as
$\rho \rightarrow \infty$, the integral along the large lower
semicircle goes to zero and we are reduced to an integral from
$-\infty$ to $+\infty$, as before. As announced above, this last
argument justifies the fact that we have extracted the Green function
from Eq.~(\ref{A1}).
Performing explicitly the integration, the Green function for topology
change reads
\begin{widetext}
\begin{eqnarray}
\label{Gi}
G_i &=& \frac{3 m y' x^{5/2}}{4\pi \hbar \kappa \mid v \mid (\chi-{\chi}')}
\exp \biggl\{\frac{i}{\hbar \kappa}\biggl[-\frac{4m(y-y')}{3(\chi-\chi ')}
+\frac{3 {\cal K}y^{1/3}(x^{4}-1)(\chi-\chi ')}{10(x^{3}+1)x}
\biggr]\biggr\} 
\exp \biggl\{\frac{1}{\kappa}\biggl[-\frac{8m(yy')^{1/2}}{3(\chi-\chi ')}
\nonumber \\
& & -\frac{3{\cal K}y^{1/3}x^{1/2}(x+1)(\chi-\chi ')}{10(x^{3}+1)}
\biggr]\biggr\}\Theta \biggl(-\frac{\cal K}{\cal K'}\biggr)\; ,
\end{eqnarray}
\end{widetext}
where the quantity $x$ has been introduced before. This is the main 
result of this section.

\par

Let us now check that the above expression possesses the correct
properties of a Green function. The Wheeler-De Witt equation for the
homogeneous background is given by
\begin{equation}
\label{hwdw}
-im\hbar\frac{\partial \Psi}{\partial\chi}  =
-\frac{3\kappa}{4} \hbar ^2 y\frac{{\partial}^2
\Psi}{\partial y ^2} - \frac{{\cal K}y^{1/3}}{2\kappa} \Psi \; .
\end{equation}
The Green function $G_i$ must satisfy the semi-classical version of
Eq. (\ref{hwdw}). Recovering the units, $G_i$ has the following form:
$G_i = C\exp(iS/\hbar)\exp(A/\hbar)$. After inserting this expression
in Eq.~(\ref{hwdw}), one obtains an equation whose real and imaginary
parts, in the limit $\hbar \to 0$, read
\begin{eqnarray}
\label{rhsc}
m\frac{\partial S}{\partial\chi}
-\frac{3\kappa}{4} y\biggl[\biggl(\frac{\partial S}{\partial y}\biggr)^2
-\biggl(\frac{\partial A}{\partial y}\biggr)^2\biggr]
-\frac{{\cal K}y^{1/3}}{2\kappa} &=& 0  \; ,
\\
\label{ihsc}
-m\frac{\partial A}{\partial\chi}
+\frac{3\kappa}{2}y\frac{\partial S}{\partial y} 
\frac{\partial A}{\partial y} &=& 0 \; .
\end{eqnarray}
Using the expression of $G_i$ derived above, see Eq.~(\ref{Gi}), one
can easily check that Eqs.~(\ref{rhsc}) and (\ref{ihsc}) are indeed
satisfied in the long-wavelength approximation. Let us notice that
terms involving $R^2=(6{\cal K}/y^{2/3})^2$ have been
neglected. Hence, $G_i$ possesses all the good properties of a Green
function signaling a topology change: it describes the transition from
a wave function describing an homogeneous spatial geometry with
positive (negative) curvature to another wave function describing
another homogeneous spatial geometry with negative (positive)
curvature [see Eq.~(\ref{green2})], and it satisfies the
semi-classical Wheeler-De Witt equation in the long-wavelength
approximation.

\subsection{Discussion}

In this section, we briefly discuss the main properties of the Green
function of Eq.~(\ref{Gi}).

\par

First of all, it is interesting to notice that, without the change of
sign obtained in the first term of the exponential, see
Eq.~(\ref{psic}), the Green function obtained through
Eq.~(\ref{green2}) would not satisfy the Wheeler-De Witt
equation~(\ref{hwdw}). This change of sign is a direct consequence of
describing the change of topology by means of complexification of
coordinates, which in turn implies an overall change of sign in the
four-metric while the transition is taking place. This confirms the
ideas of Ref.~\cite{hawking} stating that, in a topology transition,
one should pass through degenerate metrics. We notice in passing that
Eq.~(\ref{rhsc}) is not exactly the classical Hamilton-Jacobi equation
for a {\em minisuperspace} corresponding to the homogeneous metrics
considered here. There is an extra term which induces the quantum
change of topology. This is the quantum potential in the Bohm-de
Broglie interpretation language. This extra term comes from the
imaginary part $A$ of the action and indicates that such a change of
topology is in fact a quantum tunneling effect.

\par

Secondly, examining the Green function~(\ref{Gi}), one can notice that
a topology change is very improbable if $\Delta\chi\equiv\chi -\chi '$
is very short.  As $\Delta\chi$ increases, quantum changes of topology
become possible in both directions (from positive to negative
curvature and vice-versa), but when $\Delta\chi$ becomes very large,
negative to positive curvature transitions are suppressed while
positive to negative curvature transitions are enhanced. This suggests
that negative curvature topologies are preferred.  Finally, for finite
$\Delta\chi$, topology transitions between large volume spacelike
hypersurfaces are very improbable, as expected.

\section{Conclusions}

We now briefly summarize the main results reached in this article. 

\par

We have obtained a Green function describing a topology change by
using a double approximation: the semi-classical canonical quantum
gravity and the long wavelength approximation. Our considerations have
been restricted to transitions between homogeneous spacelike
hypersurfaces with different intrinsic curvatures. In this framework,
we have been able to demonstrate that transitions involving flat
hypersurfaces are forbidden. However, we have also shown that this is
no longer the case if the hypersurfaces are curved. In order to obtain
the corresponding Green functions describing transitions between
curved hypersurfaces, we have used imaginary coordinates (not only
timelike coordinate but also one spacelike coordinate) in order to
circumvent the {\em midisuperspace} problem raised in Ref.~\cite{LMPS}
(i.e. the components of the four-dimensional curvature tensor are only
functions of time). The explicit expression of the Green function has
been obtained by performing an integration in the complex plane. The
resulting Green function shows that topology changes in the direction
of negatively curved hypersurfaces are strongly enhanced as time goes
on while transitions in the opposite direction are suppressed. In a
finite amount of time, transitions between large spacelike
hypersurfaces are improbable.

\par

The formalism developed in this paper can also be applied to more
complicated models. This will be the subject of our future
investigations.

\section*{ACKNOWLEDGMENTS}

Two of us (NPN and IDS) would like to thank the Cosmology Group of
CBPF for useful discussions, and CNPq of Brazil for financial support.

\end{document}